\RequirePackage{fix-cm}
\documentclass[smallextended]{svjour3}       
\smartqed  
\usepackage{graphicx}
\usepackage{subfigure}

\newcommand{\be}[1]{\begin{equation}\label{#1}}
\newcommand{\ee}{\end{equation}}
\newcommand{\ba}[1]{\begin{eqnarray}\label{#1}}
\newcommand{\ea}{\end{eqnarray}}
\newcommand{\rf}[1]{(\ref{#1})}
\newcommand{\nn}{\nonumber}

\begin{document}

\title{Standard and helical magnetorotational instability 
}
\subtitle{How singularities create paradoxal phenomena in MHD}

\titlerunning{Standard and helical magnetorotational instability}        

\author{Oleg N. Kirillov         \and
        Frank Stefani 
}


\institute{
O.N. Kirillov \at
              Helmholtz-Zentrum Dresden-Rossendorf\\
              P.O. Box 510119, D-01314 Dresden, Germany \\
              Tel.: +49-351-2602154\\
              \email{o.kirillov@hzdr.de}          
           \and
           F. Stefani \at
              Helmholtz-Zentrum Dresden-Rossendorf\\
              P.O. Box 510119, D-01314 Dresden, Germany \\
              Tel.: +49-351-2603069\\
              \email{f.stefani@hzdr.de}          
}

\date{Received: date / Accepted: date}

\maketitle

\begin{abstract}
The magnetorotational instability (MRI) triggers turbulence
and enables outward transport of angular momentum in
hydrodynamically stable rotating shear flows, e.g., in accretion disks.
What laws of differential rotation are susceptible to the
destabilization by
axial, azimuthal, or helical magnetic field? The answer to this question, which is   
vital for astrophysical and experimental
applications, inevitably leads to the study of spectral
and geometrical singularities
on the instability threshold. The singularities provide a
connection between seemingly discontinuous
stability criteria and thus explain several paradoxes in
the theory of MRI that were poorly understood since the 1950s.

\end{abstract}

\section{Introduction}
\label{intro}

In 1890 Maurice Couette, a student of Gabriel Lippmann and Joseph Boussinesq, defended his thesis
``Etudes sur le frottement des liquides'' and was awarded his
doctorate  at the Sorbonne ``with all white balls'' and
\textit{cum laude} for the experiments with a viscometer of
his own design \cite{c94,c88,c90}. Seventy years later
Evgeny Velikhov, then a physics student of Stanislav Braginsky at the M.V. Lomonosov Moscow State University,
discovered the magnetorotational instability of the Couette-Taylor flow \cite{v59}.

The fates of the first scientific works of both young scientists were similar 
in a sense that the reaction of the scientific community
in both cases was quiescent for almost 30 years, until Geoffrey Taylor
investigated stability of the rotating Couette flow in 1923 \cite{t23}
and Steven Balbus and John Hawley demonstrated in 1991 the crucial role
of the magnetorotational instability for the explanation of transition to
turbulence and thus the anomalous viscosity in accretion disks
surrounding gravitating bodies \cite{bh91}.

\begin{figure}
  \includegraphics[width=\textwidth]{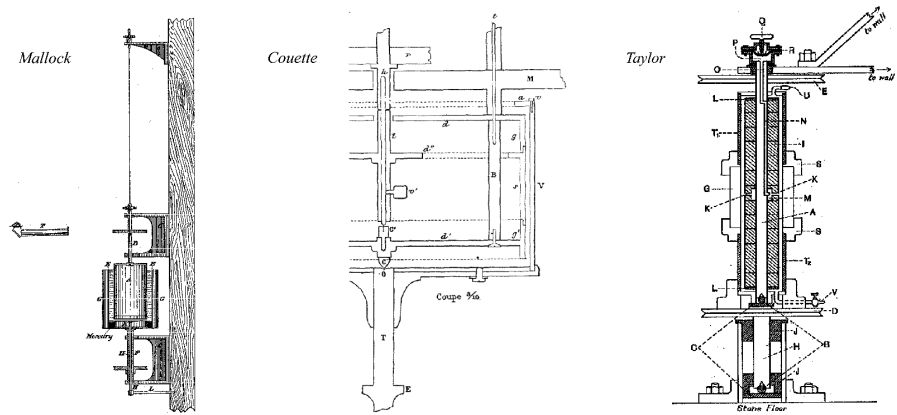}
\caption{Original drawings of the viscometers of (left) Mallock \cite{m88,m96} with either outer or inner rotating cylinder and (center)
Couette \cite{c94,c90} with rotating outer cylinder and (right) the experimental apparatus of Taylor \cite{t23} in which both cylinders could rotate.}
\label{fig:1}       
\end{figure}

The aim of Couette was to measure the kinematic viscosity of water.
In 1888 \cite{c88} he reported on the design of a viscometer that
he presented at the
1889 Universal Exhibition in Paris \cite{c94}.
In the Couette viscometer the liquid occupied a space between two
co-axial cylinders,
the outer one rotating while the inner one fixed, Fig.~\ref{fig:1}. Couette
found that at small speeds of rotation the moment of the drag which the
fluid exerted on the inner cylinder was indeed proportional to the
velocity of the outer cylinder, from which the kinematic viscosity
was determined. At higher speeds the drag increased at a greater
rate than the velocity, indicating the onset of turbulent motion.

\begin{figure}
  \includegraphics[width=\textwidth]{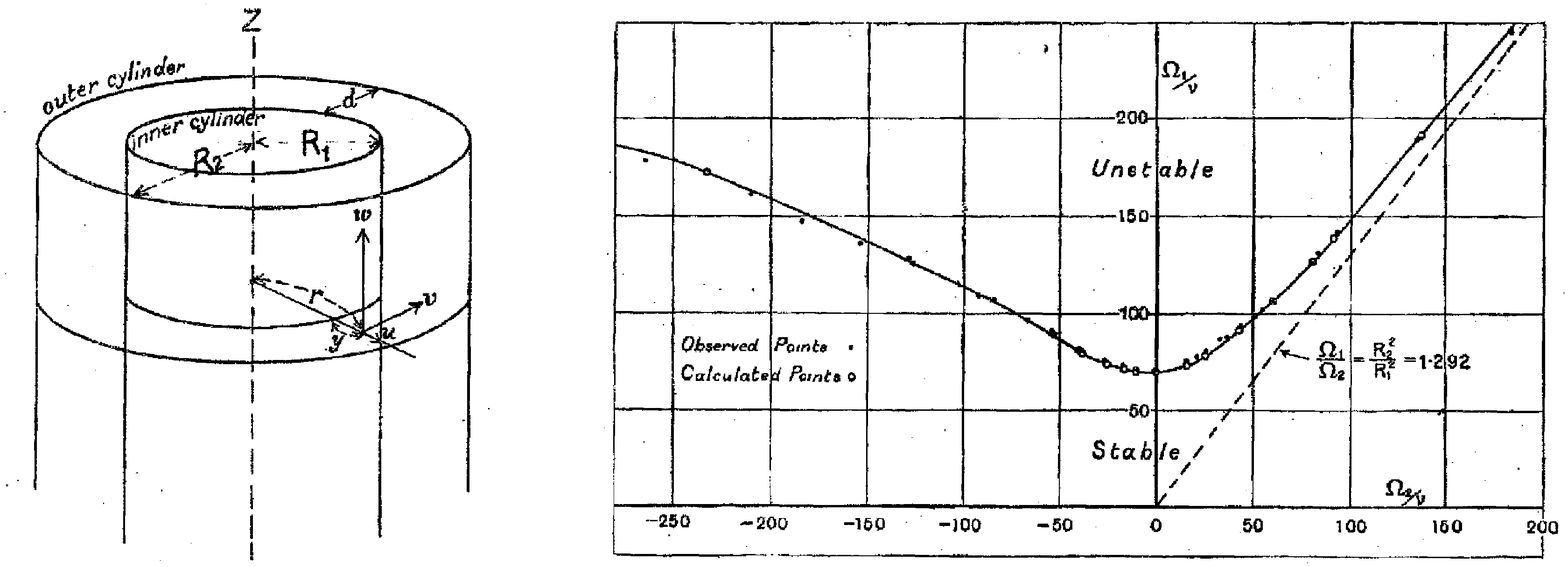}
\caption{Original drawings of the geometry of the Taylor's model and of the stability diagram in the $(\Omega_2/\nu,\Omega_1/\nu)$-plane for $R_1 = 3.55$ cm and $R_2= 4.035$ cm \cite{t23}. Dashed line is the Rayleigh's inviscid stability boundary \cite{r17}.}
\label{fig:2}       
\end{figure}

In his thesis Couette referred \cite{c94} to the work of Arnulph Mallock
from Rayleigh's laboratory \cite{m88}  who independently designed a      
similar device with either the inner or the outer cylinder rotating,  
Fig.~\ref{fig:1}. Mallock confirmed Couette's results, but in the case when
the inner cylinder rotated and the outer one not, he surprisingly observed
instability of the fluid at all speeds that he used \cite{m88,m96}. Although
the effect had been anticipated  by Stokes already in 1848 \cite{D91}, it was 
explained (in the inviscid approximation) by Rayleigh only in 1917   
\cite{r17}.
According to Rayleigh's criterion, an inviscid rotating flow is stable with respect 
to axisymmetric perturbations provided that its angular momentum increases radially
\be{i1}
\frac{1}{R^3}\frac{d}{dR}(\Omega R^2)^2>0.
\ee
When this criterion is not fulfilled, the balance between the
centrifugal force and a pressure gradient is broken and the
flow is centrifugally unstable. In particular, the inviscid
fluid between two co-rotating and co-axial cylinders of infinite
lengths and radii $R_1<R_2$ is unstable if and only if
\be{i2}
\Omega_1 R_1^2> \Omega_2 R_2^2,
\ee
where $\Omega_1$ and $\Omega_2$ are the angular velocities
of the inner and outer cylinders, respectively, see Fig.~\ref{fig:2}.

Limitations in the design of the experiments by Couette and
Mallock did not allow the full verification of the criterion \rf{i2}.   
Besides, the steady flow in their viscometers was not
close enough to two-dimensional  because of relatively     
small length-to-diameter ratio.						
This motivated Geoffrey Taylor to construct a slimmer Couette
cell making both co-rotation and counter-rotation of the cylinders    
possible, Fig.~\ref{fig:1}. In his 1923 work \cite{t23}
Taylor performed a linear stability analysis of the Navier-Stokes
equations in case of infinite length cylinders and managed         
to find a stability diagram in the $(\Omega_2/\nu,\Omega_1/\nu)$-plane,
see Fig.~\ref{fig:2}. It turned out that the viscosity, $\nu$, modifies the
Rayleigh criterion in such a manner that it becomes only a sufficient
stability condition and that the viscous stability boundary
asymptotically tends to the Rayleigh line in case of the
co-rotating cylinders, Fig.~\ref{fig:2}.
Moreover, the viscous flow is stable at small speeds of the inner
cylinder when the outer one is at rest or in motion, while it
inevitably becomes unstable when the velocity of the inner
cylinder exceeds a critical value. This is in contradiction
with the observations of Couette and Mallock that the viscous
flow becomes unstable for large velocities of the outer cylinder
when the inner does not move and is unstable at all speeds of the
inner cylinder when the outer is at rest. The latter discrepancy
is due to the fact that Mallock's lowest speed of rotation, 2 rpm,
was still larger than the critical value calculated for the size of the     
cylinders he used \cite{D91} while the former is essentially caused
by the insufficient axial elongation of the viscometers of Mallock
and Couette.

The stability boundary extracted from Taylor's experimental data
\cite{t23} perfectly agreed with that followed from his linear
stability analysis, Fig.~\ref{fig:2}.
Furthermore, Taylor's experiments revealed that with the violation
of the stability threshold, the rotating Couette flow bifurcates to a
secondary steady state characterized by counter-rotating toroidal
vortices (the Taylor vortex flow). Extending the parameters deeper inside    
the instability domain results in flows with even more complicated spatiotemporal
patterns \cite{t23,CI94}. Therefore, the instability of the
Couette-Taylor (CT) flow is analogous to the static (divergence)
instability in structural mechanics \cite{SH06}.

The excellent correspondence that Taylor obtained between theory and experiment
demonstrated the correctness of the Navier-Stokes equations and of the
no-slip boundary condition for the fluid at the cylinder walls \cite{D91}
as well as it proved the applicability of linear stability analysis to the
CT-flow.
After the influential work \cite{t23}, Couette-Taylor cells     
became a standard equipment for laboratory testing
hydrodynamical and magnetohydrodynamical theories.

In 1953 Chandrasekhar first considered
the CT-flow of a weakly electrically conducting viscous fluid in the presence
of the uniform magnetic field that is parallel to the axis of rotation of
the cylinders \cite{Ch53}.
He demonstrated that in this case characterized by the very small ratio
of the kinematic viscosity coefficient, $\nu$, to the magnetic diffusivity
coefficient, $\eta$, i.e. by the magnetic Prandtl
number ${\rm Pm:=\nu\eta^{-1}}\ll 1$, the magnetic field stabilizes the
hydrodynamically unstable CT-flow \cite{Ch53}. The article of Chandrasekhar
has been brought to the attention of Evgeny Velikhov by his supervisor
Stanislav Braginsky who posed a problem
on the influence of the axial magnetic field on the hydrodynamically
stable CT-flow---the very question that had not been addressed in \cite{Ch53}.   

\begin{figure}
  \includegraphics[width=0.95\textwidth]{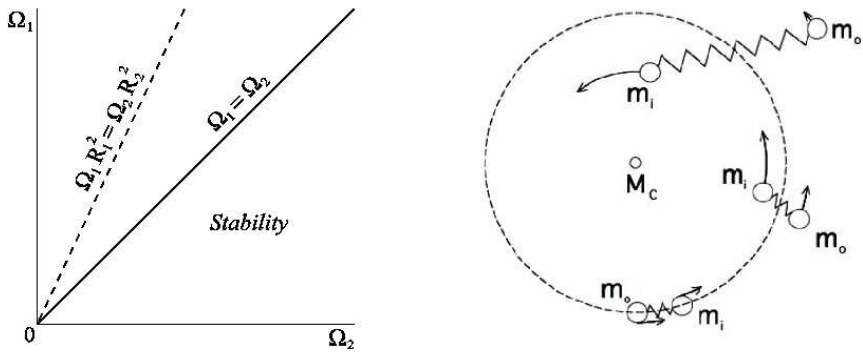}
\caption{(left) A diagram of stability with respect to axisymmetric perturbations of the Couette-Taylor flow of ideally electrically conducting inviscid fluid with an axial magnetic field applied \cite{v59}; dashed line is  Rayleigh's inviscid stability boundary \cite{r17}. (right) A paradigmatic mechanism of magnetorotational instability in accretion disks \cite{B09}. The action of the magnetic field is equivalent to that
of elastic springs that lead to the exchange of angular momentum between the fast inner mass $m_i$ and the slow outer mass $m_o$.}
\label{fig:3}       
\end{figure}

In contrast to Chandrasekhar, Velikhov assumed that the liquid is
both inviscid and perfectly conducting.                                 
In 1959 he established a new sufficient criterion of stability with
respect to the axisymmetric perturbations in the form
\be{i3}
\frac{d\Omega^2}{dR}>0,
\ee
or, in terms of the angular velocities of the cylinders,
\be{i4}
\Omega_2>\Omega_1,
\ee
see the left panel of Fig.~\ref{fig:3}.
In a subsequent work of 1960 Chandrasekhar confirmed this result \cite{Ch60}.

Both Velikhov and Chandrasekhar pointed out that the new stability conditions  
\rf{i3} and \rf{i4} do not depend on the magnetic field strength $B$ which implies   
that in the limit $B\rightarrow0$ they
do not converge to the Rayleigh's criteria \rf{i1} and \rf{i2}
valid for $B=0$, as is illustrated by the left panel of Fig.~\ref{fig:3}.
In presence of dissipation the convergence is possible \cite{Ji01,RZ01,WB02,D05,AH73}.
This dependence of the
instability threshold on the sequence of taking the two
limits of vanishing magnetic field and
vanishing electrical resistivity constitutes the famous
\textit{Velikhov-Chandrasekhar paradox}.
Its physical explanation has been given in 				
terms of Alfv\'{e}n's theorem \cite{Al42} that in a fluid             
of zero resistivity the
magnetic field lines are frozen-in to the fluid,
independent on the strength of the magnetic
field \cite{v59,Ch60}.

Even half a century after the publications of Velikhov and Chandrasekhar
the `dubious' jump   
in the threshold of the \textit{magnetorotational instability} (MRI)
has not been fully understood and the paradox remained unresolved
\cite{AH73,B03,Tassoul04,v05,s09,GKP10} despite ``much of the fluid and stellar
community was aware of the instability, however, and of its curious behavior
of ostensibly changing the Rayleigh criterion discontinuously'' \cite{B03}.
Maybe this is not so surprising when taking into account 			
that even the astrophysical relevance of MRI to destabilize a differentially    
rotating flow in accretion disks around gravitating celestial bodies         
remained underappreciated during a long period until it                        
was rehabilitated by Balbus and Hawley in 1991.					

The central problem here is that accretion disks typically    
rotate according to Kepler's law, $\Omega(R) \sim R^{-3/2}$
which results in an angular momentum $R^2\Omega(R)\sim R^{1/2}$
that fulfills Rayleigh's stability criterion \rf{i1}. Such stable, non-turbulent
disks would not allow the outward directed angular momentum
transport that is necessary for the infalling disk matter to accrete
into the central object. In their seminal paper \cite{bh91},
Balbus and Hawley had highlighted
the key role of the MRI in this process by showing that a weak,
externally applied
magnetic field is a trigger for the instability that
actually taps into the rotational energy of the flow.

In the perfectly conducting fluid, the magnetic field
lines are `frozen' into it
tethering fluid elements like a spring \cite{B03,BH98},
see Fig.~\ref{fig:3}(right). If such a couple is perturbed,
the magnetic `tether' retards the faster inner element that
has to move to the lower Keplerian orbit and simultaneously it
accelerates the slower outer fluid element that thus has to
move to the higher orbit. The separation between the elements grows
with time yielding instability. Remarkably, this simple mechanical
analogue of MRI proposed by Balbus and Hawley \cite{bh91,B03,BH98},
is a working principle of numerous engineering projects developed
since 1960s that involve
momentum exchange tethers for the orbital transfer of satellites
\cite{bl93}. Well-known is instability of the orbiting
ring of connected satellites as well as of the orbiting
flexible and extensible ring in the context of studies of
formation of planetary rings \cite{B81,BL85}.

In the reference frame comoving with a small patch of the
magnetized accretion disk and rotating at the angular velocity    
$\Omega_0=\Omega(R_0)$,
the leading order WKB equations governing the
evolution of its local radial ($x$) and azimuthal ($y$)
displacements in the vicinity of a fiducial point with
the radius $R_0$, are \cite{B03}
\ba{i5}
\ddot x - 2\Omega_0 \dot y + \left(R_0\left.\frac{d\Omega^2}{dR}\right|_{R=R_0}+
\omega_A^2\right)x&=&0,\nn\\
\ddot y + 2\Omega_0 \dot x + \omega_A^2 y &=&0,
\ea
where the Alfv\'en frequency, $\omega_A$, measures the intensity of the
magnetic tension force.
With $\omega_A=0$ and Keplerian rotation $\Omega(R)\sim R^{-3/2}$ the
equations \rf{i5} are reduced to
the Hill-Clohessy-Wiltshire ones \cite{h78,cw60} that describe in
particular the relative motion of two satellites.

Writing down the
characteristic equation of system \rf{i5} and taking into account that at the
onset of standard MRI with only an axial magnetic field applied (which is a non-oscillatory instability)
the critical eigenvalue is vanishing \cite{h04}, we get the instability threshold
(see also \cite{Ch60})
\be{i7}
{\rm Ro}:= \frac{1}{2}\frac{R}{\Omega} \frac{d\Omega}{dR}=
- \frac{\omega_A^2}{4\Omega_0^2},
\ee
where $\rm Ro$ is the Rossby number (evaluated at $R=R_{0}$ in \rf{i7}) that
indicates the deviation of the rotating  shear flow from the solid body
rotation, ${\rm Ro}=0$.
The latter is the threshold resulting from the sufficient stability criterion
of Velikhov-Chandrasekhar (${\rm Ro}>0$).
The actual threshold of MRI given by equation \rf{i7} depends on the
magnetic field strength through $\omega_A$ and
therefore it coincides with the criterion \rf{i3} in the limit $\omega_A=0$
which deviates from the non-magnetic value ${\rm Ro}=-1$ following from the
Rayleigh criterion \rf{i1}. This non-uniqueness of the critical Rossby number
in the non-magnetic limit is another manifestation of the Velikhov-Chandrasekhar paradox.

In contrast to the non-magnetic Couette-Taylor case, the theory of MRI
was ahead of laboratory experiments.          			
First interesting experimental results were obtained only in
2004 in a spherical Couette flow of liquid sodium \cite{s04}.
In this experiment, the authors observed correlated modes of velocity and
magnetic field perturbation in parameter regions
which are quite typical for MRI. However, the background state
in this spherical Couette experiment
was already fully turbulent, so that the
original goal to show the basic destabilizing effect of a magnetic
field was not met. Recent works \cite{Ho04,GJG11} have also shown that 
the observed effects might be alternatively explained in terms 
of two different sorts of non-axisymmetric 
magnetic instabilities in spherical Couette flow.       

At Princeton University, work is going on to identify MRI
in a CT-experiment with liquid gallium, and first encouraging
results, including the observation of nonaxisymmetric
magneto-Coriolis (MC) waves, have been recently reported \cite{n10,Ji10}.
The Princeton facility had been designed to investigate the standard		
version of MRI (SMRI) with only a vertical magnetic field
being applied. SMRI is known to work only with magnetic Reynolds numbers 
(Rm) in the order of 1 or larger. Rm is proportional to
the hydrodynamic Reynolds number
according to Rm = PmRe, where Pm is the magnetic Prandtl number.
For liquid metals Pm is typically in the range
$10^{-6}-10^{-5}$. Therefore, in order to achieve Rm $\sim$ 1, we need
${\rm Re}=10^5-10^6$, and wall-constrained flows (in contrast to wall-free
Keplerian flows) with such high Re are usually turbulent,
whatever the linear stability analysis might tell \cite{pl11}.
This is the point which makes SMRI experiments,
and their interpretation, so cumbersome \cite{b11}.

This situation changed drastically when Hollerbach and
R\"udiger considered the effect of adding an azimuthal magnetic field
to the axial one \cite{HR05}. Indeed, it was shown
\cite{HR05,R10} that the resulting helical
MRI (HMRI), as we now call it, is then possible at far smaller
Reynolds numbers and magnetic field amplitudes than SMRI,
making HMRI an ideal playground for liquid metal experiments.

First experimental evidence for HMRI was obtained in 2006
at the liquid metal facility PROMISE (Potsdam ROssendorf
Magnetic InStability Experiment) which is basically a CT-cell
made of concentric rotating copper walls, filled with GaInSn
(a eutectic which is liquid at room temperatures).
In \cite{S06,R06,S08a} it was shown that the HMRI traveling  
wave appears only in the predicted finite window of the magnetic
field intensity, with a frequency of the traveling wave that was
in rather good accordance with numerical simulations.            
Some disturbing effects of this early version (PROMISE 1), connected     
with the recirculating  radial jet at midheight of the cylinder,       
were overcome in the follow-up PROMISE 2 experiment            
by splitting the axial end caps to suppress the Ekman          
pumping  \cite{S08,S09}. By comparing experimental and numerical (based on
\cite{PRIEDE}) results for   
a wide variety of parameter dependencies, it was possible to identify    
the observed instability as an absolute one, distinguishing it clearly   
from  a noise triggered convective instability as speculated on
in \cite{LIU}.   

Despite SMRI being a                                     
non-oscillatory instability and HMRI being an oscillatory one,   
there is a continuous and monotonic transition between them
when Re and the magnetic field strength are
increased simultaneously \cite{HR05,R08}. This is all the more    
remarkable in that HMRI has been identified with
the destabilization of an inertial wave in apparent contrast to SMRI that
is a destabilized slow Magneto-Coriolis wave \cite{LGHJ06,LV07,PGG07,ks10}. 
The transition from SMRI to
HMRI, which  are characterized by substantially different scaling laws,
involves the origination of a spectral exceptional point
\cite{skm05} and a transfer of instability between the modes \cite{ks10}.

It is remarkable, too, that even in the limit of vanishing electrical
conductivity (${\rm Pm} \rightarrow 0$),
the helical magnetic field is able to trigger an instability
although the instantaneous growth of the energy of any
perturbation must be smaller than in the field-free case---the
{\textit{paradox of inductionless HMRI}} \cite{PGG07}.
In this inductionless case, however, the local WKB analysis
in the small-gap approximation prohibits helical magnetorotational instability
when ${\rm Ro}$ exceeds the Liu limit of $2-2\sqrt{2}\simeq -0.828$ \cite{LGHJ06}.
Thus the inductionless HMRI works only for comparably steep rotation profiles (i.e.,
slightly above the Rayleigh line of ${\rm Ro}=-1$) and disappears for profiles as
flat as the Keplerian one with ${\rm Ro}=-0.75$, see also \cite{LV07,R08}.
This behaviour has been experimentally confirmed  in the PROMISE experiment \cite{s09}.

Applicability of HMRI to higher Rossby numbers is
sensitive also to
electrical boundary conditions \cite{rh07}.
In addition to this,
HMRI at the Rossby numbers slightly above the Liu limit was
observed already   
in the WKB approximation at small but finite Pm in \cite{ks10}.

The ultimate upper limit of the critical
Ro in this case is an
intriguing question, in particular because of
new arguments that arose recently
from investigations of the saturation regime of MRI. For
the case of small
magnetic Prandtl numbers
(as they are typical for the outer parts of accretion disks    
\cite{BaHe08,LeLo07}),
Umurhan \cite{u10} speculated about a saturated
rotation profile with regions of reduced shear, sandwiched by
regions of strengthened shear. For those latter
regions with steeper than Keplerian profiles,
HMRI could indeed become of significant relevance.

Despite a more than a century-long history,     
hydrodynamic and hydromagnetic stability of rotating
shear flows remains
a vibrant area of research, full of intriguing paradoxes
and mathematical, computational, and experimental challenges.
Below extending the recent works of the authors \cite{ks10,ks11} we present a viewpoint that relates some of the mentioned
effects to singularity theory---an approach that had already
proven its efficiency in the field of dissipation-induced
instabilities \cite{Bo56,Ar71,Le80,Le82,GKL90,BKMR94,HR95,L03,K04,K07,KM07,KV10}.

\section{Mathematical setting}
\label{sec:1}
The standard set of non-linear equations of
dissipative incompressible magnetohydrodynamics
\cite{LV07,R08,ks10} consists of the Navier-Stokes equation for
the fluid velocity ${\bf u}$
\be{m1}
\frac{\partial {\bf u}}{\partial t}+({\bf u} \cdot \nabla)
{\bf u} = - \frac{1}{\rho} \nabla \left(p+
\frac{{\bf B}^2}{2\mu_0}\right)+
\frac{1}{\mu_0 \rho}({\bf B}\cdot \nabla){\bf B}+\nu \nabla^2 {\bf u},
\ee
and of the induction equation for the magnetic field ${\bf B}$
\be{m2}
\frac{\partial {\bf B}}{\partial t}=\nabla \times ({\bf u}
\times {\bf B}) + \eta \bigtriangledown^2 {\bf B},
\ee
where $p$ is the pressure, $\rho=const$
the density, $\nu=const$ the kinematic
viscosity,
$\eta=(\mu_0 \sigma)^{-1}$ the magnetic diffusivity, $\sigma$ the conductivity of the fluid,
and $\mu_0$
the magnetic permeability of free space.
Additionally, the mass continuity equation for incompressible flows
and the solenoidal condition for the magnetic induction yield
\be{m3}
\nabla \cdot {\bf u} = 0,\quad  \nabla \cdot {\bf B}=0.
\ee

We consider the rotational fluid flow in the gap between the radii
$R_1$ and $R_2>R_1$, with an imposed magnetic field sustained by
currents external to the fluid.
Introducing the cylindrical coordinates $(R, \phi, z)$ we consider
the stability of a steady-state background liquid flow with           
the angular
velocity profile $\Omega(R)$ in a helical background magnetic field (a
magnetized CT-flow)
\be{m4} {\bf u}_0=R\,\Omega(R)\,{\bf
e}_{\phi},\quad p=p_0(R), \quad {\bf B}_0=B_{\phi}^0(R){\bf
e}_{\phi}+B_z^0 {\bf e}_z,
\ee
with the azimuthal component
\be{m4a}
B_{\phi}^0(R)=\frac{\mu_0 I}{2 \pi R},                                    
\ee which can be thought as
being produced by an axial current $I$. The angular velocity profile      
of the background CT-flow is \cite{s09}
\be{m4b}
\Omega(R)=\frac{\Omega_1R_1^2-\Omega_2R_2^2}{R_1^2-R_2^2}+\frac{1}{R^2}\frac{(\Omega_2-\Omega_1)R_1^2R_2^2}{R_1^2-R_2^2}.
\ee
The centrifugal acceleration of the
background flow \rf{m4b} is compensated by the pressure gradient
\be{m4d} \frac{1}{\rho}\frac{\partial
p_0}{\partial R}=R\Omega^2. \ee

\subsection{Linearization with respect to axisymmetric perturbations}
\label{sec:2}

Throughout the paper we will restrict our interest to axisymmetric
perturbations ${\bf u}'={\bf u}'(R,z)$, ${\bf B}'={\bf B}'(R,z)$,
and ${p}'={p}'(R,z)$ about the stationary solution \rf{m4}-\rf{m4b}.
Non-axisymmetric perturbations become important for the so-called azimuthal
MRI (AMRI) and the Tayler instability \cite{RHGS07,RSG11}.

With the notation
\be{l0}
\partial_t=\frac{\partial}{\partial t},\quad \partial_R=\frac{\partial}{\partial R},\quad \partial_z=\frac{\partial}{\partial z}, \quad \partial_R^{\dagger} = \partial_R +\frac{1}{R},\quad D=\partial_R\partial_R^{\dagger}+\partial_z^2
\ee
we write the linearized equations that couple $u_R'$, $u_{\phi}'$ and $B_R'$, $B_{\phi}'$ \cite{ks10}
\be{al10}
\partial_t \tilde{E} \xi'=\tilde{H}\xi',
\ee
where                  $\xi'=(u_R',u_{\phi}',B_R',B_{\phi}')^T$, and
\be{al12}
\tilde{E}=\left(                                    
    \begin{array}{cccc}
      D & 0 & 0 & 0 \\
      0 & 1 & 0 & 0 \\
      0 & 0 & 1 & 0 \\
      0 & 0 & 0 & 1 \\
    \end{array}
  \right), \quad
\tilde{H}=\left(                               
    \begin{array}{cccc}
      \nu D^2\quad & 2\Omega \partial_z^2\quad & \frac{B_z^0}{\mu_0 \rho} D{\partial_z}\quad & -\frac{2B_{\phi}^0}{\mu_0 \rho R} \partial_z^2 \\
      -{2\Omega}(1+\rm Ro) \quad & \nu D\quad & 0\quad & \frac{B_z^0}{\mu_0\rho} {\partial_z} \\
      B_z^0 \partial_z\quad & 0\quad & \eta D\quad & 0 \\
      \frac{2B_{\phi}^0}{R}\quad & B_z^0 \partial_z \quad& {2\Omega}\rm Ro \quad & \eta D \\
    \end{array}
  \right).
\ee
The resulting multiparameter family of operator matrices equipped with boundary conditions can be investigated by numerical
or perturbative \cite{K09} methods. In the following we use the local WKB approximation.

\subsection{Local WKB approximation}

We expand all the background quantities in Taylor series around a fiducial point $(R_0,z_0)$ and retain only
the zeroth order in terms of the local coordinates $\tilde R=R-R_0$ and $\tilde z=z-z_0$
to obtain the operator matrix equation
\ba{w0}
\partial_t \tilde{E}_0 \xi'=\tilde{H}_0\xi'                       
\ea
with
\be{w1}
\tilde{E}_0=\left(                                                
    \begin{array}{cccc}
      D_0 & 0 & 0 & 0 \\
      0 & 1 & 0 & 0 \\
      0 & 0 & 1 & 0 \\
      0 & 0 & 0 & 1 \\
    \end{array}
  \right), ~~
\tilde{H}_0=\left(                                           
    \begin{array}{cccc}
      \nu ({D_0})^2 ~~& 2\Omega_0 \partial_{\tilde z}^2 ~~& \frac{B_z^0}{\mu_0 \rho} D_0{\partial_{\tilde z}} ~~&
      -\frac{2B_{\phi}^0}{\mu_0 \rho R_0} \partial_{\tilde z}^2 \\
      -{2\Omega_0}(1+\rm Ro) ~~& \nu D_0 ~~& 0 ~~& \frac{B_z^0}{\mu_0\rho} {\partial_{\tilde z}} \\
      B_z^0 \partial_{\tilde z}~~ & 0 ~~& \eta D_0 ~~& 0 \\
      \frac{2B_{\phi}^0}{R_0} ~~& B_z^0 \partial_{\tilde z} ~~& {2\Omega_0}\rm Ro ~~& \eta D_0 \\
    \end{array}
  \right),
\ee
where
\be{w2}
\Omega_0=\Omega(R_0),\quad \quad B_{\phi}^0=B_{\phi}^0(R_0), \quad
D_0=\partial_{\tilde R}^2+\partial_{\tilde z}^2+\frac{\partial_{\tilde R}}{R_0}-\frac{1}{R_0^2}.
\ee

Equation \rf{w0} is a linear PDE with the constant coefficients in the local variables $(\tilde R, \tilde z)$
for the perturbed quantities $\xi '$, which is valid as long as  $\tilde R$ and $\tilde z$ are
small in comparison with the characteristic radial and vertical length scales (the so-called narrow-gap approximation \cite{DR81}).
A plane wave solution to the equation \rf{w0} is
\ba{w3}
{\xi}'&=&\hat{\xi}\exp{( i k_R \tilde R + i k_z \tilde z)},\quad
\hat\xi=\tilde{\xi}\exp{(\gamma t)},
\ea
where $\tilde{\xi}$ is a vector of constant coefficients and $\hat\xi=(\hat u_{R}, \hat u_{\phi}, \hat B_{R}, \hat B_{\phi})^{T}$.

In the WKB approximation we restrict the analysis to the short-wave modes with the wave numbers satisfying $k_R \gg \frac{1}{R_0} $ which allows us to
neglect the terms $\frac{ik_R}{R_0}-\frac{1}{R_0^2}$ in \rf{w0}. In view of this, substituting \rf{w3} into
equation \rf{w0}, yields the leading order WKB equations that describe the onset of instability of a
CT-flow with a helical external magnetic field
\be{e7}
\dot \xi = H \xi,\quad H{=}\left(
      \begin{array}{cccc}
        -\omega_{\nu} & 2\Omega_0 \alpha^2 & {i\omega_{A}} & -2\omega_{A_{\phi}}{\alpha^2} \\
        -2\Omega_0(1{+}{\rm Ro})  & -\omega_{\nu} & 0 & {i\omega_{A}} \\
        {i\omega_{A}} & 0 & -\omega_{\eta} & 0 \\
        2\omega_{A_{\phi}} & {i\omega_{A}} & 2\Omega_0{\rm Ro} & -\omega_{\eta} \\
      \end{array}
    \right),
\ee
where $\xi=(\hat u_R,\hat u_{\phi}, \hat B_R(\mu_0\rho)^{-1/2}, \hat B_{\phi}(\mu_0\rho)^{-1/2})^T$, $\alpha=k_z/k$, the total wave number is defined through $k^2=k_z^2+k_R^2$,
$\omega_{\nu}=\nu k^2$ and $\omega_{\eta}=\eta k^2$ are the viscous and resistive frequencies, and
 the Alfv\'en frequencies of the axial and azimuthal magnetic field components are
\be{w6b}
\omega_A^2=\frac{k_z^2 (B_z^0)^2}{\mu_0 \rho},\quad
\omega_{A_{\phi}}^2=\frac{(B_{\phi}^0)^2}{\mu_0\rho R_0^2},
\ee
respectively \cite{ks10}.

\section{Stability analysis}

The stability of solutions to the equation \rf{e7} is
determined by the roots $\gamma$ of the dispersion equation
\be{d1}
P(\gamma)=\gamma^4+a_1\gamma^3+a_2\gamma^2+(a_3+ib_3)\gamma+a_4+ib_4=0,
\ee
where $P(\gamma)=\det(H-\gamma E)$ and $E$ is the unit matrix.
We write the coefficients of the complex polynomial \rf{d1} in the form \cite{R08,LGHJ06,LV07,ks10}
\ba{d2}
a_1&=&2(\omega_{\nu}+\omega_{\eta}),\nn\\
a_2&=&(\omega_{\nu}+\omega_{\eta})^2+2(\omega_A^2+\omega_{\nu}\omega_{\eta})+4\alpha^2\Omega_0^2(1+{\rm Ro})+4\alpha^2 \omega_{A_{\phi}}^2,\nn\\
a_3&=&2(\omega_{\eta}+\omega_{\nu})(\omega_A^2+\omega_{\eta}\omega_{\nu})+
8\alpha^2\Omega_0^2(1+{\rm Ro})\omega_{\eta}+4\alpha^2(\omega_{\eta}+\omega_{\nu})
\omega^2_{A_{\phi}},\nn\\
a_4&=&(\omega_A^2+\omega_{\nu}\omega_{\eta})^2-4\alpha^2\omega_A^2\Omega_0^2+4\alpha^2\Omega_0^2(1+{\rm Ro})(\omega_A^2+\omega_{\eta}^2)+
4\alpha^2\omega_{\nu}\omega_{\eta}\omega_{A_{\phi}}^2,\nn\\
b_3&=&-8\alpha^2\Omega_0\omega_A\omega_{A_{\phi}},\nn\\
b_4&=&-4\alpha^2\Omega_0\omega_A\omega_{A_{\phi}}(2\omega_{\eta}+\omega_{\nu})-4\alpha^2 \Omega_0(1+{\rm Ro}) \omega_A\omega_{A_{\phi}}(\omega_{\eta}-\omega_{\nu}).
\ea

With $\omega_{A_{\phi}}=0$ the coefficients $b_3$ and $b_4$ vanish while the others simplify.
Composing the Hurwitz matrix of the resulting real polynomial,
we write the Lienard and Chipart criterion of asymptotic stability \cite{M66}
\be{d3}
 a_4>0,\quad  a_2>0,\quad h_1= a_1>0,\quad h_3= a_1 a_2 a_3- a_1^2 a_4- a_3^2>0.
\ee
Under the physical assumption that $\omega_{\nu}\ge0$ and $\omega_{\eta}\ge0$, the last two inequalities are automatically satisfied,
because
\ba{d4}
h_3&=&
4\omega_A^2(\omega_{\eta}+\omega_{\nu})^2((\omega_{\eta}+\omega_{\nu})^2+
4\Omega_0^2\alpha^2)\nn\\&+&
4\omega_{\eta}\omega_{\nu}((\omega_{\eta}+\omega_{\nu})^2+4\Omega_0^2\alpha^2(1+{\rm Ro}))^2>0.
\ea
On the other hand, condition $a_4>0$ implies that
\be{d5}
{\rm Ro}>{\rm Ro}_{c}:=-\frac{(\omega_A^2+
\omega_{\nu}\omega_{\eta})^2+4\Omega_0^2\omega_{\eta}^2\alpha^2}
{4\Omega_0^2\alpha^2(\omega_A^2+\omega_{\eta}^2)}>-1-\frac{\omega_{\nu}^2}{4\alpha^2\Omega_0^2},
\ee
which yields $a_2>0$.
Therefore, the four stability conditions \rf{d3} are reduced to
the only \rf{d5} that yields the critical Rossby number
(${\rm Ro}_{c}$) of SMRI (above which the flow is stable).

\begin{figure}
  \includegraphics[width=0.99\textwidth]{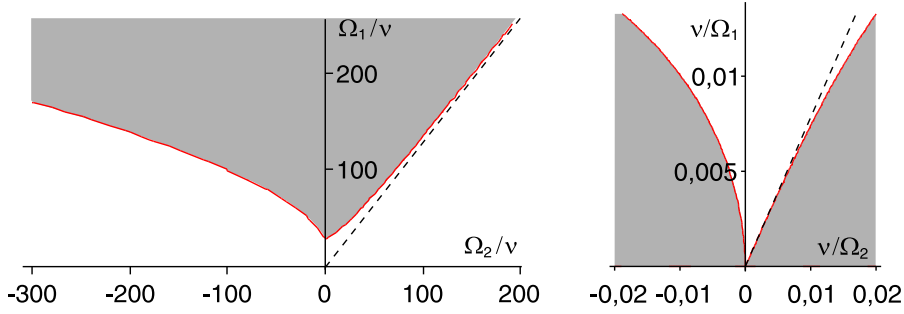}
\caption{Stability diagrams according to the WKB thresholds \rf{d14a} and \rf{d14b} with centrifugal instability shown in grey and stability in white calculated
for  the Taylor values $R_1=3.55$ cm and $R_2=4.035$ cm: (left) in $({\Omega}_2/\nu,{\Omega}_1/\nu)$-plane \cite{EY95} and (right) in
$(\nu/{\Omega}_2,\nu/{\Omega}_1)$-plane;
dashed line is the Rayleigh threshold.}
\label{fig:4}       
\end{figure}

Alternatively, this threshold follows from equations \rf{e7} that can be rewritten
as a non-conservative gyroscopic system
\be{d8}
\ddot
u +
(D +\Omega_0(1+\alpha^2){J})
\dot
u +
(N +K)
u=0,
\ee
where  $u=(\hat u_R, \hat u_{\phi})^T$, ${N}=\Omega_0(\omega_{\eta}(1+\alpha^2)+{\rm Ro}(\omega_{\eta}-\omega_{\nu})){J}$,
\be{d9}
{K}= \left(
            \begin{array}{ll}
              \omega_A^2+\omega_{\nu}\omega_{\eta} & k_{12} \\
              k_{12} & \omega_A^2+\omega_{\nu}\omega_{\eta}+4\alpha^2\Omega_0^2{\rm Ro} \\
            \end{array}
          \right)
\ee
with $k_{12}=\Omega_0(\omega_{\eta}(1-\alpha^2)+{\rm Ro}(\omega_{\eta}-\omega_{\nu}))$, and
\be{d10}
{J}=\left(
                                                                                                             \begin{array}{rr}
                                                                                                               0 & -1 \\
                                                                                                              1 & 0 \\
                                                                                                             \end{array}
                                                                                                           \right),~~
{ D}=\left(
          \begin{array}{cc}
            \omega_{\nu}+\omega_{\eta} & \Omega_0(1-\alpha^2) \\
            \Omega_0(1-\alpha^2) & \omega_{\nu}+\omega_{\eta} \\
          \end{array}
        \right).
\ee
For $\alpha=1$, $\omega_{\nu}=0$, and $\omega_{\eta}=0$, Eq. \rf{d8} is similar to
the model \rf{i5} and has the same dispersion equation.

Stable perturbations  have $\Re \, \gamma \le 0$ provided that
$\gamma$ with $\Re \, \gamma = 0$ is a semi-simple eigenvalue of the eigenvalue problem
corresponding to \rf{d8}. The growing solutions of SMRI are non-oscillatory with $\Im \gamma=0$. Therefore, $\gamma=0$ implies that $\det({N}+{K})=0$
at the threshold of SMRI  which gives the critical Rossby number, ${\rm Ro}_{c}$.

\subsection{Non-magnetic Couette-Taylor flow}

In the absence of the magnetic field $\omega_{A}=0$ and
\be{d11}
{\rm Ro}_{c}=-1-\frac{\omega_{\nu}^2}{4\alpha^2\Omega_0^2},
\ee
which exactly reproduces the result of Eckhardt and Yao obtained by means of the geometrical optics stability analysis \cite{EY95}.

Following \cite{EY95}, we write the condition for centrifugal instability as
\be{d11a}
{\rm Ro}_{c}<-1-\frac{{\nu}^2}{4\Omega_0^2}\frac{ (1+k_z^2/k_R^2)^3}{k_z^2/k_R^2}k_R^4\le -1-\frac{{\nu}^2}{4\Omega_0^2}\frac{ 27}{4}k_R^4.
\ee
Using the definition of the Rossby number \rf{i7} and the profile of angular velocity \rf{m4b} in the inequality
\rf{d11a}, we transform it into
\be{d12a}
\frac{\Omega_1R_1^2-\Omega_2R_2^2}{R_1^2-R_2^2}\Omega(R_1)\le
\frac{\Omega_1R_1^2-\Omega_2R_2^2}{R_1^2-R_2^2}\Omega(R_0)\le -\nu^2\frac{27}{16}k_R^4,
\ee
where $\Omega(R_1)=\Omega_1$. Restricting the radial wave number from below further by $|k_R|>\pi/(R_2-R_1)$ in case of co-rotating cylinders ($\Omega_2\ge 0$)
and by $|k_R|>\pi/(R_c-R_1)$ in case of counter-rotation ($\Omega_2< 0$), where $R_c$ is the radius at which $\Omega(R)$ changes its sign, see \cite{EY95}
\be{d13a}
R_c=\sqrt{\frac{\Omega_1-\Omega_2}{\Omega_1R_1^2/R_2^2-\Omega_2}}R_1,
\ee
and taking into account the narrow-gap approximation, equivalent to the condition $(R_2-R_1)/R_1 \ll 1$,
we derive the WKB approximations to the Taylor instability domain for $\Omega_2\ge 0$
\be{d14a}
\Omega_2<\frac{R_1^2}{R_2^2}\left(1-\frac{\Omega_c^2}{\Omega_1^2} \right)\Omega_1
\ee
and for $\Omega_2< 0$
\be{d14b}
\Omega_2>\frac{R_1^2}{R_2^2}\left(1-\frac{\Omega_1^2}{\Omega_c^2} \right)\Omega_1,
\ee
where
\be{d15a}
\Omega_c=\nu\sqrt{\frac{27\pi^4}{8R_1(R_2-R_1)^3}}.
\ee

In Fig.~\ref{fig:4}(left) we see that the local WKB approximation \rf{d11} qualitatively and quantitatively correctly reproduces the Taylor's stability diagram that was obtained from the analysis of the global boundary eigenvalue problem. Since the modern experiments intending to observe SMRI require very high Reynolds numbers \cite{s04,pl11,b11}, it is instructive to redraw this stability diagram in the $(\nu/{\Omega}_2,\nu/{\Omega}_1)$-plane, Fig.~\ref{fig:4}(right). In this plane the stability domain shown in white in Fig.~\ref{fig:4}(right) has a self-intersection singularity at the origin, which at least illustrates the recent experimental evidence that at high Reynolds numbers the ratio ${\Omega}_{1}/{\Omega}_{2}$ is `the dominant control
parameter for rotating shear flows' \cite{pl11}. The existence of such a singularity that causes high sensitivity of the instability threshold to the variation of the parameters might be a reason for a controversy in the experiments \cite{s04,pl11} that indicate both laminar and turbulent states of the CT-flow at high Reynolds numbers \cite{b11}.

                \begin{figure}[tp]
    \centering
    \includegraphics[angle=0, width=0.95\textwidth]{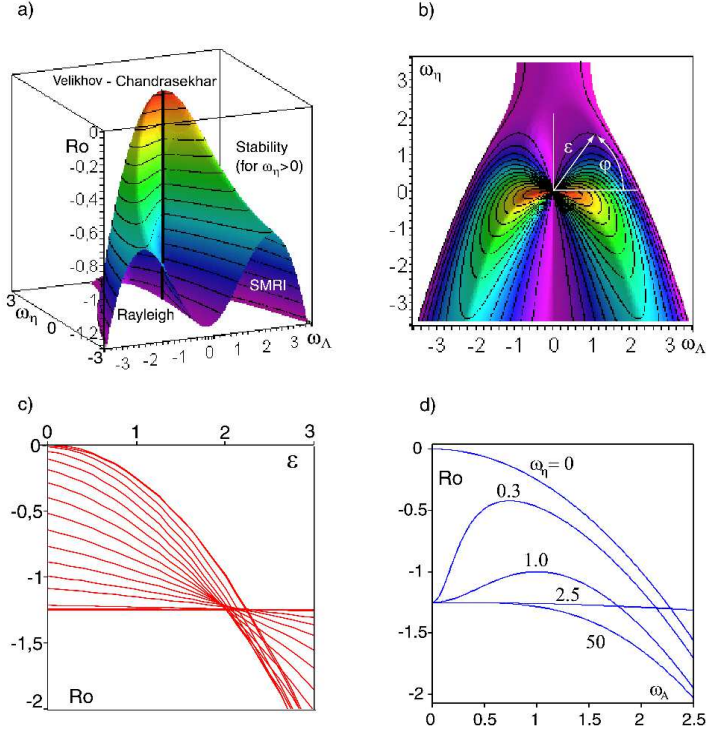}
    \caption{(a) The critical Rossby number of SMRI as a function of  $\omega_A \sim {\rm Lu}{\rm Pm}^{-1}$ and $\omega_{\eta} \sim {\rm Pm}^{-1}$ for $\omega_{\nu}=1$, $\alpha=1$, $\Omega_0=1$, i.e. for  ${\rm Re}=1$. (b)
 Top view of the surface. (c) Cross-sections of the surface along the rays specified by the Lundquist number, or, equivalently, by the angle $\varphi$ that varies from $0$ to $1.5$ through the equal intervals $\Delta \varphi =0.1$; the horizontal line corresponds to $\varphi=\pi/2$. (d) Transition between the case of low (Velikhov 1959) and high (Chandrasekhar 1953) resistivity. }
    \label{fig:5}
    \end{figure}

\subsection{Velikhov-Chandrasekhar paradox in standard MRI}

The Velikhov-Chandrasekhar paradox occurs at infinite ${\rm Pm}=\omega_{\nu}\omega_{\eta}^{-1}$ and means that in
the ideal MHD case ($\omega_{\eta}=0$, $\omega_{\nu}=0$) the limit $\omega_{A}\rightarrow0$
yields Velikhov's value ${\rm Ro}_{\rm c}=0$ as the instability threshold
rather than Rayleigh's limit ${\rm Ro}_{\rm c}=-1$ of the non-magnetic case $(\omega_{A}=0$, $\omega_{\nu}=0)$.

With  $\omega_A=\varepsilon\cos\varphi$ and $\omega_{\eta}=\varepsilon\sin\varphi$ in \rf{d5}, we obtain
\be{e5}
{\rm Ro}_{\rm c}= -\frac{(\varepsilon\cos^2\varphi+\omega_{\nu}\sin\varphi)^2+4\alpha^2\Omega_0^2\sin^2\varphi}{4\alpha^2\Omega_0^2},
\ee
which for $\varepsilon \rightarrow 0$ reduces to
\be{e6}
{\rm Ro}_{\rm c}= -\left(1+\frac{1}{4{{\rm Re}}^2}\right)\sin^2\varphi=-\frac{1+(2{\rm Re})^{-2}}{1+{{\rm Lu}}^2},
\ee
where ${\rm Re}=\Omega_0 \alpha \omega_{\nu}^{-1}$  and ${\rm Lu}=\omega_A\omega_{\eta}^{-1}$ are the Reynolds and the Lundquist numbers, respectively.
Introducing the new parameter ${\rm Ro}'=(1+4{{\rm Re}}^2(1+2{\rm Ro}))(1+4{{\rm Re}}^2)^{-1}$ we find that in the
$(\omega_A,\omega_{\eta},{\rm Ro}')$-space Eq. \rf{e6} defines a so-called
{\it ruled surface} $(\varepsilon,\varphi)\mapsto (\varepsilon\cos\varphi,\varepsilon\sin\varphi,\cos n\varphi)$ with $n=2$, which is a canonical equation for the Pl\"ucker conoid of degree $n=2$ \cite{BG88,KH10}. The surface
according to Eq. \rf{d5} tends to the Pl\"ucker conoid when $\varepsilon=\sqrt{\omega_A^2+\omega_{\eta}^2}\rightarrow 0$.

This surface is shown in the $(\omega_A,\omega_{\eta},{\rm Ro})$-space in Fig.~\ref{fig:5}(a) and in projection to the $(\omega_A,\omega_{\eta})$-plane in Fig.~\ref{fig:5}(b) for ${\rm Re}=1$. For each $\alpha$, $\omega_{\nu}$, and $\Omega_0$ it has the same Pl\"ucker conoid singularity, i.e. an
interval of self-intersection along the ${\rm Ro}$-axis and two Whitney umbrella
singular points at its ends. This singular structure implies non-uniqueness for the
critical Rossby number when simultaneously $\omega_A=0$ and $\omega_{\eta}=0$.

Indeed, for a given $\rm Lu$, tending the magnetic field to zero along
a ray $\omega_A=\omega_{\eta}{\rm Lu}$ in the $(\omega_A,\omega_{\eta})$-plane
results in a value of the Rossby number specified by Eq. \rf{e6},
see Fig.~\ref{fig:5}(c). The limit value of the critical Rossby number
oscillates between the ideal MHD value ${\rm Ro}_{\rm c}=0$
for ${\rm Lu}=\infty$ $(\varphi=0)$
and the non-magnetic (Taylor) value
${\rm Ro}_{\rm c}=-1-(2{\rm Re})^{-2}$ for
${\rm Lu}=0$ $(\varphi=\pi/2)$,
which resolves the Velikhov-Chandrasekhar paradox.

Physically, the Lundquist number determines the `lifetime' of the magnetic field line that is frozen into the fluid.
In the ideal MHD case ${\rm Lu}=\infty$ means that the field does not diffuse from the fluid. At the lower values of
${\rm Lu}$ resistivity destroys the magnetic tension effect which prevents $\rm Ro$ from reaching the solid body rotation value in the
limit of vanishing magnetic field.

Fig.~\ref{fig:5}(d) demonstrates transition between the cases of high conductivity (Velikhov 1959)
and of low conductivity (Chandrasekhar 1953) separated by the threshold
$
\omega_{\eta}= ({\omega_{\nu}^2+4\Omega_0^2\alpha^2})(2\omega_{\nu})^{-1}.
$ In the latter case the axial magnetic field stabilizes the hydrodynamically unstable CT-flow.
Fig.~\ref{fig:5}(d) also illustrates the conclusions of Acheson and Hide that in the presence of small but finite resistivity in the limit of vanishing $\omega_A$ ``the stability or otherwise of the system will then be decided essentially by Rayleigh's criterion" \cite{AH73}.

\begin{figure}[tp]
 \centering
  \includegraphics[width=0.99\textwidth]{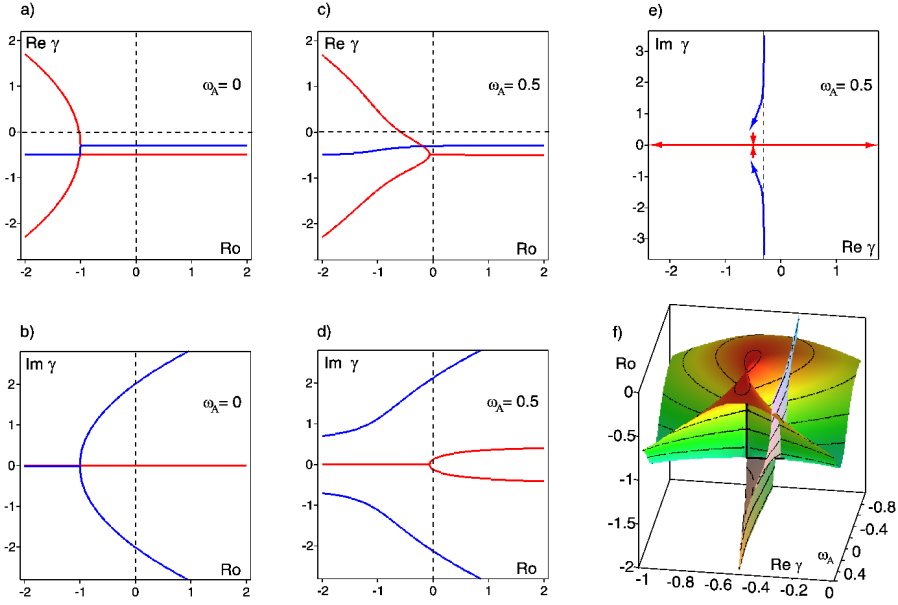}
\caption{Growth rates and frequencies of the perturbation for  $\Omega_0=1$, $\alpha=1$, $\omega_{\nu}=0.3$, $\omega_{\eta}=0.5$ and (a,b) $\omega_A=0$ and (c,d,e) $\omega_A=0.5$; (f) growth rates surfaces in the $(\Re\gamma, \omega_A, {\rm Ro})$-space.}
\label{fig:4a}       
\end{figure}

What eigenvalue behavior corresponds to the singular threshold of SMRI?
In the absence of the magnetic field the roots of the dispersion equation \rf{d1} are exactly
\be{e7a}
\gamma_{1,2}=-\omega_{\nu}\pm i2\alpha\Omega_0\sqrt{{\rm Ro}+1}, \quad \gamma_{3,4}=-\omega_{\eta}.
\ee
The first two roots bifurcate and one of them becomes positive at the critical Rossby number given by equation \rf{d11}, see Fig.~\ref{fig:4a}(a,b,e).
The blue curves there represent the inertial waves whose interaction yields the centrifugal instability in the inviscid case \cite{S33}.
The Rayleigh line is thus characterized by the double zero eigenvalue with the Jordan block. Viscosity shifts this doublet
to the left part of the complex plane \cite{GG93}.
This scenario corresponds to the lower Whitney umbrella singularity on the SMRI threshold surface
shown in Fig.~\ref{fig:5}(a).

With $\omega_A \ne 0$ the merging of the inertial waves becomes imperfect, see Fig.~\ref{fig:4a}(c,d),
while the formerly damped roots $\gamma_{3,4}$ experience a bifurcation at the Rossby number that is close to the Velikhov-Chandrasekhar value.
In Fig.~\ref{fig:4a}(d) the red branches correspond to the slow Magneto-Coriolis (MC) waves while the blue ones --- to the fast Magneto-Coriolis waves. Bifurcation of the slow MC-waves precedes the onset of SMRI (equation \rf{d5}) with the decrease of $\rm Ro$.  In the absence of viscosity and resistivity the roots of the dispersion equation \rf{d1} corresponding to slow- and fast MC-waves are exactly
\be{e8}
\gamma^2=-2\Omega_0^2\alpha^2(1+{\rm Ro})-\omega_A^2 \pm 2\Omega_0\alpha\sqrt{\Omega_0^2\alpha^2(1+{\rm Ro})^2+\omega_A^2}.
\ee
The corresponding double zero eigenvalue at $\omega_A=0$ and $\rm Ro=0$ is related to the upper Whitney umbrella singularity at the threshold surface of SMRI in Fig.~\ref{fig:5}(a).

Transition between these two bifurcations happens in the presence of resistivity and viscosity and is described by means of the
slices of two singular eigenvalue surfaces shown in Fig.~\ref{fig:4a}(f). The surface corresponding to the roots $\gamma_{1,2}$ is locally equivalent to the Pl\"ucker conoid of degree $n=2$ while that of the roots $\gamma_{3,4}$ is locally equivalent to the Pl\"ucker conoid of degree $n=1$ \cite{BG88}.

\subsection{Paradox of inductionless helical magnetorotational instability }

Now we turn over to the paradox of inductionless HMRI which is
related to a similar geometric singularity as discussed above.

After scaling the spectral parameter as $\gamma=\lambda \sqrt{\omega_{\nu}\omega_{\eta}}$, we express the appropriately normalized coefficients \rf{d2} by means of the dimensionless Rossby number $({\rm Ro})$, magnetic Prandtl number $({\rm Pm})$, helicity parameter $\beta=\alpha \omega_{A_{\phi}}\omega_A^{-1}$ of the external magnetic field, Hartmann (${\rm Ha}={\rm Lu}{\rm Pm}^{-1/2}$), and Reynolds $({\rm Re})$ numbers. Additional transformations yield the coefficients of the                        
dispersion equation $P(\lambda)=0$
\ba{e8a}
a_1&=&2(1 +{\rm Pm}^{-1})\sqrt{\rm Pm},\nn\\
a_2&=&2({1+(1+2{\beta}^2){\rm Ha}}^2)+4{\rm Re}^2   (1+{\rm Ro}){\rm Pm}+a_1^2/4, \nn \\
a_3&=&a_1(1+(1+2{\beta}^2){{\rm Ha}}^2)+8{{\rm Re}}^2(1+{\rm Ro})\sqrt{\rm Pm},\nn \\
a_4
&=&\left(1{+}{{\rm Ha}}^2\right)^2+4{\beta}^2{{\rm Ha}}^2
+4{\rm Re}^2(1+{\rm Ro}({\rm Pm}{{\rm Ha}}^2+1)),\nn \\
b_3&=&-8\beta{{\rm Ha}}^2{\rm Re}\sqrt{\rm Pm},~~
b_4=b_3(1+(1-{\rm Pm}){\rm Ro}/2)/\sqrt{\rm Pm}.
\ea

        \begin{figure}
    \begin{center}
    \includegraphics[angle=0, width=0.9\textwidth]{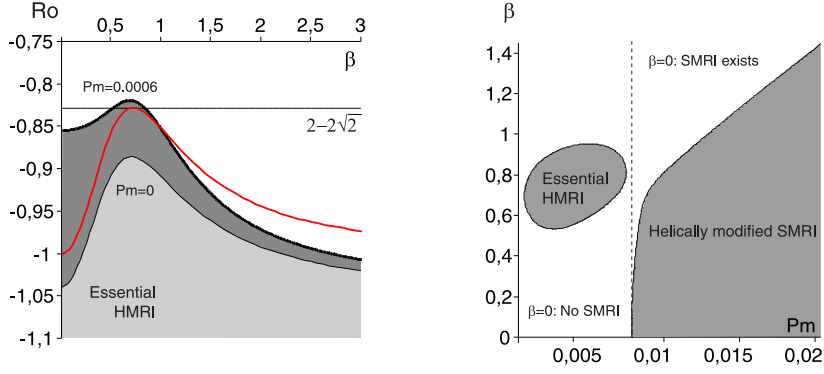}
    \end{center}
     \caption{(left) In the inductionless limit ${\rm Pm}=0$ the instability domain (light grey, shown for ${\rm Ha}=19$ and ${\rm Re}=900$) is always under the majorating red curve given by Eq.~\rf{r2} that touches the Liu limit ${\rm Ro}=2-2\sqrt{2}$ at $\beta=\sqrt{2}/2$; for  ${\rm Pm}\ne 0$ however this is no longer true and the instability domain (dark grey, shown for ${\rm Ha}=19$ and ${\rm Re}=900$) can partly lie above the Liu limit. (right) HMRI island (shown for ${\rm Ha}=5$, ${\rm Re}=100$, and ${\rm Ro}=-0.85$) in the $({\rm Pm},\beta)$-plane exists at such low ${\rm Pm}$ at $\beta \ne 0$ where SMRI at $\beta = 0$ doesn't \cite{ks10}.}
    \label{fig:6}
    \end{figure}

The analogue of the Routh-Hurwitz
conditions for the complex polynomials---the Bilharz criterion \cite{Bi44}---
requires positiveness of all diagonal even-ordered minors of the so-called
Bilharz matrix composed of the coefficients \rf{e8a}
\ba{s3}
m_1&=&a_3a_4+b_3b_4>0,\quad
m_2=(a_2a_3-a_1a_4)m_1-a_2^2b_4^2>0,\nn \\
m_3&=&(a_1a_2-a_3)m_2-(a_1^2a_4a_2+(a_1b_3-b_4)^2)m_1\nn\\
&+&a_1a_4(b_4a_2(2b_4-a_1b_3)+a_1^2a_4^2)>0,\nn \\
m_4&=&a_1m_3-a_1a_3m_2+ (a_3^3+a_1^2b_4b_3-2a_1b_4^2)m_1\nn\\
&+&a_1b_4^2a_4(a_1a_2 -a_3)-b_4^2a_3^2a_2+b_4^4>0.
\ea
When the last of the stability conditions \rf{s3} is fulfilled,
the remaining inequalities are satisfied automatically \cite{ks10}. Therefore,
the threshold of HMRI is defined by the equation $m_4(\beta,{\rm Re},{\rm Ha},{\rm Pm},{\rm Ro})=0$.
For $\beta=0$ the dispersion equation and thus the threshold for HMRI reduce to that of
SMRI \cite{ks10}.

        \begin{figure}
    \begin{center}
    \includegraphics[angle=0, width=0.42\textwidth]{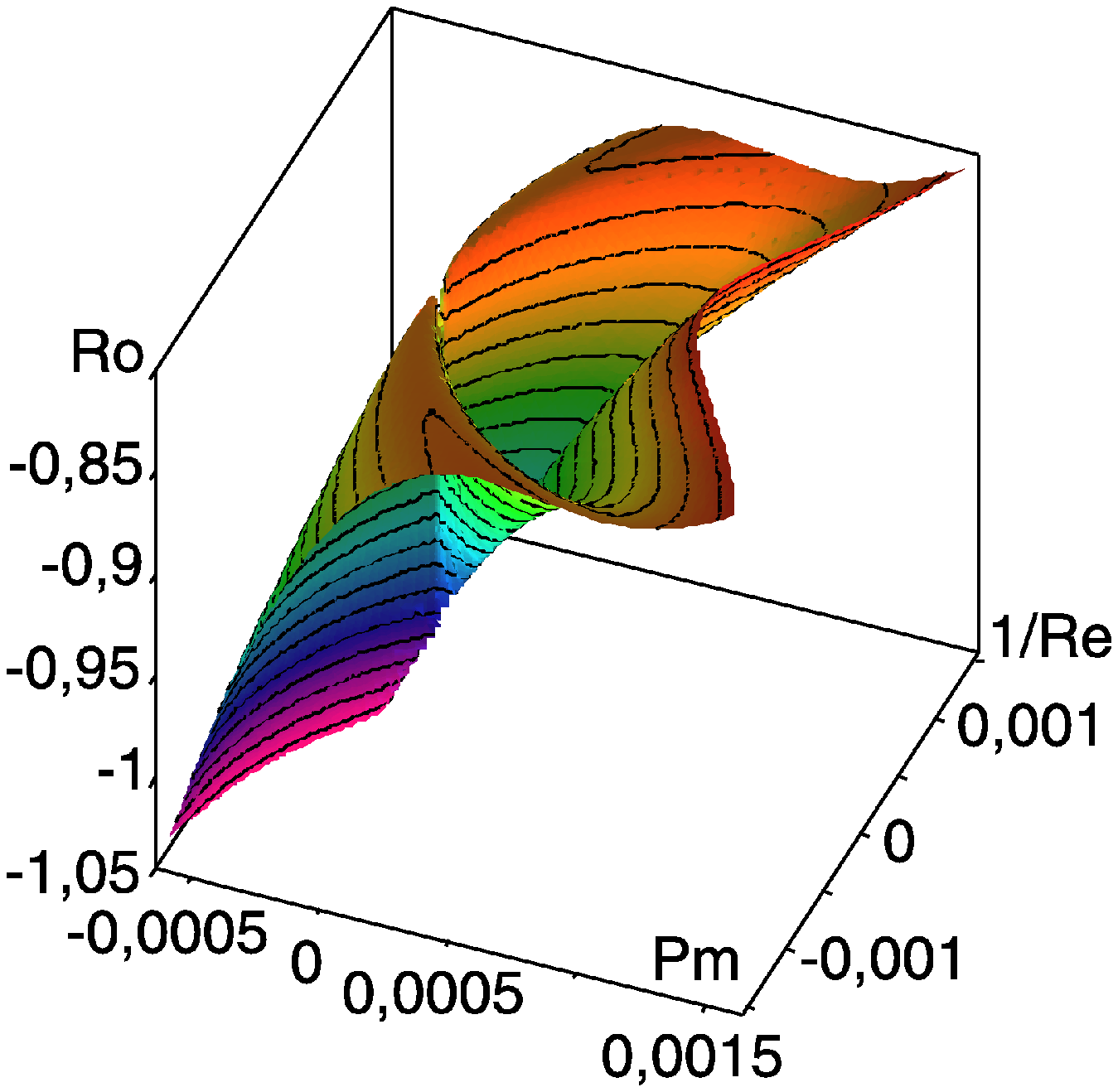}
    \hspace{.2in}
    \includegraphics[angle=0, width=0.42\textwidth]{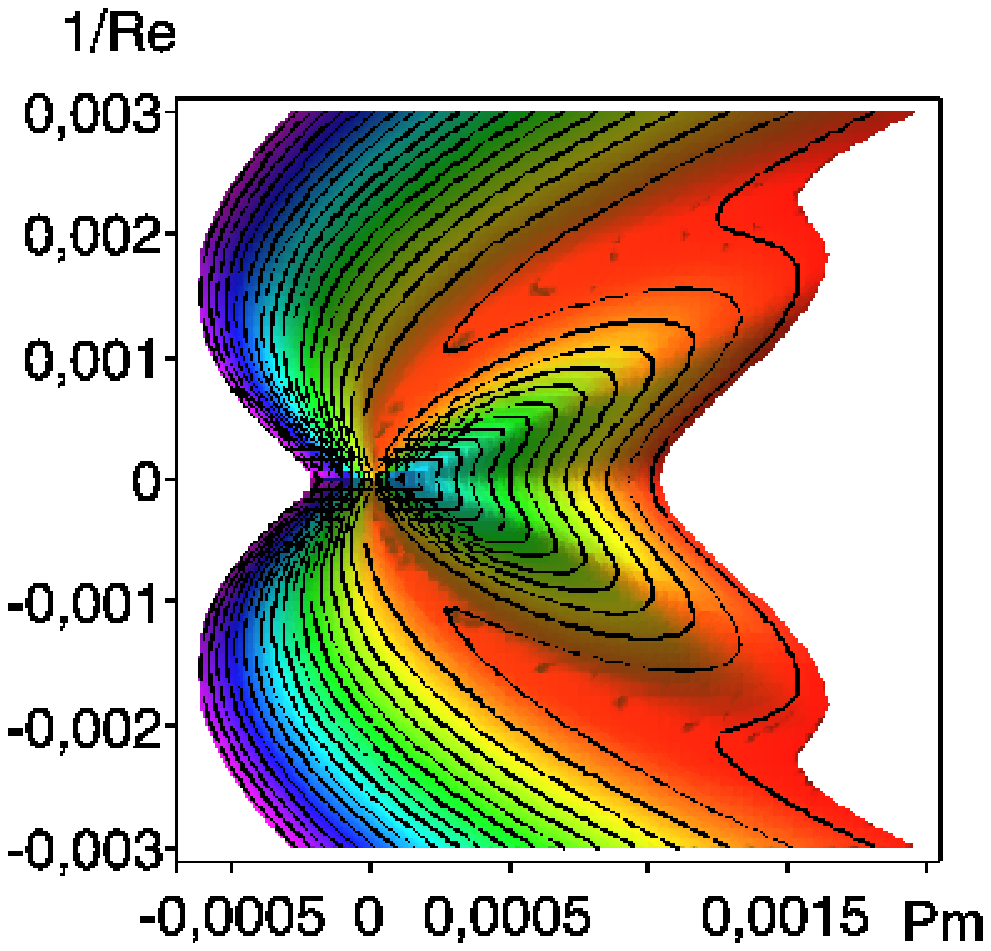}
    \end{center}
    \caption{Instability threshold in the presence of the helical magnetic field  for ${{\rm Ha}}=15$ and $\beta=0.7$ in the $({\rm Pm},{{\rm Re}}^{-1},{\rm Ro})$-space and in projection to the $({\rm Pm},{{\rm Re}}^{-1})$-plane.}
    \label{fig:7}
    \end{figure}

In the inductionless limit ${\rm Pm} \rightarrow 0$ the critical Rossby number for the onset of HMRI follows from the equation $m_4=0$
in the explicit form
\ba{i1a}
{\rm Ro}&=& \frac{\left(1{+}{\rm Ha}^2\right)^2{+}4{\beta}^2{\rm Ha}^2(1{+}{\beta}^2 {\rm Ha}^2)}{2{\rm Ha}^4{\beta}^2}
-\frac{2{\beta}^2 {\rm Ha}^2{+}{\rm Ha}^2{+}1}{2{\rm Ha}^4{\beta}^2}\\
&\times&
\sqrt{\left(1{+}{\rm Ha}^2\right)^2{+}4{\beta}^2{\rm Ha}^2(1{+}{\beta}^2{\rm Ha}^2){+}
\frac{{\rm Ha}^4{\beta}^2}{{\rm Re}^2}\left(\left(1{+}{\rm Ha}^2\right)^2{+}4{\beta}^2{\rm Ha}^2\right)}.\nn
\ea
In the limit ${\rm Re}\rightarrow\infty$ and ${\rm Ha}\rightarrow\infty$ this critical value is majorated by
\be{r2}
{\rm Ro}(\beta)=
\frac{1+4{\beta}^4-(1+2{\beta}^2)\sqrt{1+4{\beta}^4}}{2{\beta}^2},
\ee
with the maximum at the well-known Liu
limit ${\rm Ro}_{\rm c}=2-2\sqrt{2}\simeq-0.828$ when $\beta=\sqrt{2}/2\simeq0.707$ \cite{LGHJ06,ks10}.
The line \rf{r2} is shown red in Fig.~\ref{fig:6}(left).
Nevertheless, at small but finite $\rm Pm$, the HMRI domain shown in dark grey in Fig.~\ref{fig:6}(left)
can exceed the Liu limit for some choice of $\rm Ha$ and $\rm Re$.

To understand how far beyond the Liu limit HMRI can exist, we show  in Fig.~\ref{fig:7}
a typical critical
surface $m_4=0$ in the $({\rm Pm},{{
\rm Re}}^{-1},{\rm Ro})$-space for the special
parameter choice ${\rm Ha}=15$ and $\beta=0.7$.
On the $\rm Ro$-axis we find a self-intersection and
two Whitney umbrella
singularities at its ends. At the upper singular point, i.e.
exactly at ${\rm Pm}=0$, the critical Rossby number is given by Eq.~\rf{i1a} in the limit
${\rm Re} \rightarrow \infty$.

In Fig.~\ref{fig:7} we see that the case with ${\rm Pm}=0$ is
connected to the case ${\rm Pm} \ne 0$
by the Pl\"ucker conoid singularity, quite similar as it was
discussed
for the paradox of Velikhov and Chandrasekhar.
Interestingly, ${\rm Ro}_{\rm c}$ for the onset of
HMRI can indeed
increase when ${\rm Pm}$ departs from zero which happens
along curved pockets of HMRI.
The two side bumps of the curve ${\rm Re}^{-1}({\rm Pm})$ in a
horizontal slice of the surface correspond to the domains
of the {\it essential HMRI} while the central hill
marks the {\it helically modified SMRI} domain, according to
the classification introduced in \cite{ks10}, see also Fig.~\ref{fig:6}(right).
For small $\rm Pm$ the essential HMRI occurs at higher
${\rm Ro}$ than the helically modified SMRI, while for
some finite value of ${\rm Pm}$
the central hill and the side bumps
get the same value of $\rm Ro_c$.
Most remarkably, there is a value of $\rm Ro_c$
at which the two side bumps of
the curve ${\rm Re}^{-1}({\rm Pm})$ disappear completely.
This is the
maximal possible value for the essential HMRI,
at least at the
given $\beta$ and ${\rm Ha}$. Now we can ask: how does this
limit behave if we send ${\rm Ha}$ to infinity, and to which value
of ${\rm Lu}$ does this correspond?

\begin{figure}
    \begin{center}
    \includegraphics[angle=0, width=0.9\textwidth]{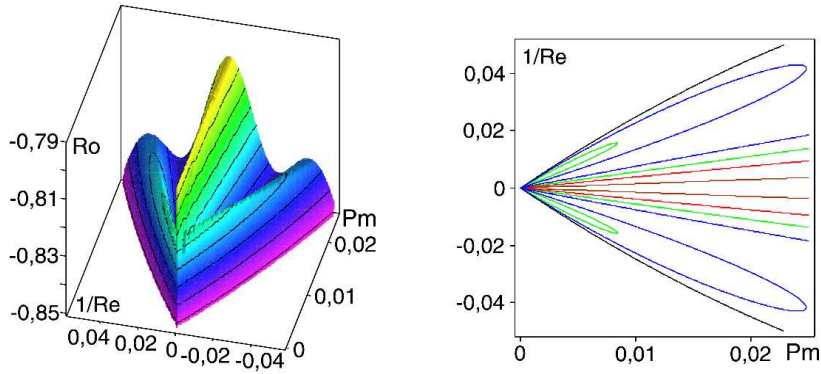}
    \end{center}
    \caption{ (left) The critical Rossby number for ${{\rm Lu}}=0.5$ and $\beta=0.6$
    in the $({\rm Pm},{{\rm Re}}^{-1},{\rm Ro})$-space and (right) its cross-sections in the
$({\rm Pm},{{\rm Re}}^{-1})$-plane for (black) {\rm Ro}=-0.842, (blue) {\rm Ro}=
-0.832, (green) {\rm Ro}=-0.822, (red) {\rm Ro}=-0.812, (brown) {\rm Ro}=-0.802 \cite{ks11}.}
    \label{fig:8}
    \end{figure}

Actually, with the increase in $\rm Ha$ the stability boundary
preserves its shape and simultaneously it
compresses in the direction of zero $\rm Pm$.
Substituting ${\rm Ha}={\rm Lu} {\rm Pm}^{-1/2}$ into the
equations \rf{e8a}, we plot again the surface
$m_4=0$ in the $({\rm Pm},{{\rm Re}}^{-1},{\rm Ro})$-space, but now
for a given $\beta$ and $\rm Lu$, Fig.~\ref{fig:8}(left).

The corresponding cross-sections of the instability
domain in the $({\rm Re}^{-1},{\rm Pm})$-plane are shown in
Fig.~\ref{fig:8}(right).
At a given value of ${\rm Ro}$ there exist three domains
of instability with the boundaries shown in blue and
green. Two sub-domains that have a form of a petal correspond to the
HMRI. They are
bounded by closed curves with a
self-intersection singularity
at the origin.
They are also elongated in a preferred direction that in
the $({\rm Re}^{-1},{\rm Pm})$-plane corresponds to a
limited range of the magnetic Reynolds number
${\rm Rm}={\rm Pm}{\rm Re}$.
The central domain, which corresponds to the
helically modified SMRI, has
a similar singularity at the origin and is unbounded
in the positive ${\rm Pm}$-direction.
In comparison with the central domain where ${\rm Rm}>1$, within the side petals
${\rm Rm}<1$.

Now we reconsider again the limit ${\rm Pm} \rightarrow 0$, while keeping
$\rm Lu$ as a free parameter.
At the origin all the boundaries of the petals can be
approximated by the straight lines
$
{\rm Pm}={\rm Rm}{{\rm Re}}^{-1}.
$
Substituting this expression into equation $m_4=0$, we find that
the only term that does not depend on ${\rm Pm}$ is a polynomial
$
Q({\rm Rm},{\rm Lu},\beta,{\rm Ro})=p_0+p_1{\rm Rm}^2+p_2 {\rm Rm}^4+p_3{\rm Rm}^6,
$
with the coefficients
\ba{sq5}
p_0&=&{\rm Lu}^4(4{\beta}^4{\rm Lu}^2+2{\beta}^2+4{\rm Lu}^2{\beta}^2+1)^2\nn\\
p_1&=&4(-{\beta}^2(1+20{\rm Lu}^4{\beta}^2+2{\rm Lu}^4+8{\beta}^2{\rm Lu}^6\nn\\
&+&16{\beta}^6{\rm Lu}^6+24{\rm Lu}^6{\beta}^4+
4{\rm Lu}^2+8{\beta}^2{\rm Lu}^2+20{\beta}^4{\rm Lu}^4){\rm Ro}^2\nn\\
&+&(16{\rm Lu}^6{\beta}^4{+}16{\beta}^6{\rm Lu}^2{+}{\rm Lu}^2{+}4{\beta}^4{+}16{\beta}^8{\rm Lu}^4{+}1{-}16{\beta}^8{\rm Lu}^6
{+}4{\rm Lu}^2{\beta}^4){\rm Ro}\nn\\
&+&1-8{\rm Lu}^2{\beta}^2({\rm Lu}^4{\beta}^2-{\beta}^2+{\rm Lu}^4-{\beta}^2{\rm Lu}^2+{\rm Lu}^2)\nn \\
&+&16{\beta}^6{\rm Lu}^2(1+
{\beta}^2{\rm Lu}^2+{\rm Lu}^2+{\rm Lu}^4)+4{\beta}^4+2{\rm Lu}^4)\nn\\
p_2&=&16({\rm Lu}^4{\beta}^4{\rm Ro}^4-{\beta}^2(-2+4{\beta}^4{\rm Lu}^4-3{\rm Lu}^2+4{\rm Lu}^4{\beta}^2){\rm Ro}^3\nn\\
&+&2{\beta}^2(3{+}4{\beta}^2{+}6{\beta}^4{\rm Lu}^4{+}4{\rm Lu}^2{+}16{\beta}^2{\rm Lu}^2{+}3{\rm Lu}^4{+}8{\rm Lu}^2{\beta}^4{+}
12{\rm Lu}^4{\beta}^2){\rm Ro}^2\nn\\
&+&(32{\beta}^4{\rm Lu}^4+16{\beta}^4+40{\rm Lu}^2{\beta}^4+2+2{\rm Lu}^2+4{\beta}^2+32{\beta}^6{\rm Lu}^4+32{\beta}^6{\rm Lu}^2)
{\rm Ro}\nn\\
&+&2+4{\rm Lu}^4{\beta}^2+8{\rm Lu}^2{\beta}^4+16{\beta}^6{\rm Lu}^2+8{\beta}^4+16{\beta}^6{\rm Lu}^4+{\rm Lu}^4+
4{\beta}^4{\rm Lu}^4)\nn\\
p_3&=&64((2{\rm Ro}{\beta}^2{+}1)^2{+}8{\rm Ro}{\beta}^4{+}4{\beta}^4{+}3{\rm Ro}^2{\beta}^2{-}{\rm Ro}^3{\beta}^2)({\rm Ro}{+}{\rm Ro}{\rm Lu}^2{+}1).
\ea

The roots of the polynomial are coefficients ${\rm Rm}$ of the linear
approximation to the instability domains at the origin in the
$({\rm Re}^{-1},{\rm Pm})$-plane. Simple roots mean
non-degenerate self-intersection of the stability boundary
at the origin. Double roots correspond to a degeneration
of the angle of the self-intersection when it collapses to
zero which happens only at the maximal critical Rossby number,
Fig.~\ref{fig:8}(left). In the $({\rm Lu}, \beta, {\rm Ro})$-space
a set of points that correspond to multiple roots of the
polynomial $Q$ is given by the discriminant surface $64\Delta^2p_0p_3=0$,  where
\be{s6}
\Delta({\rm Lu},\beta,{\rm Ro}):=18p_0p_1p_2p_3-4p_1^3p_3+p_1^2p_2^2-4p_0p_2^3-27p_0^2p_3^2.
\ee

\begin{figure}
    \begin{center}
    \includegraphics[angle=0, width=\textwidth]{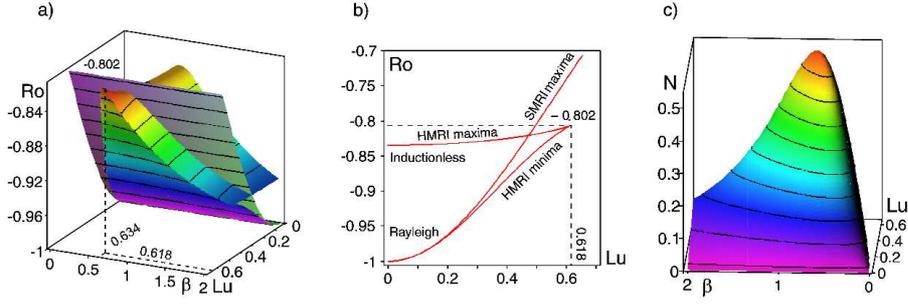}
    \end{center}
    \caption{ (a) Discriminant surface in the $({\rm Lu},\beta,{\rm Ro})$-space and (b) its cross-section \cite{ks11} at $\beta=0.634$.
    (c) Interaction parameter ${\rm N}={\rm Lu}^2 {\rm Rm}^{-1}$ at the essential HMRI maxima.}
    \label{fig:9}
    \end{figure}

The surface $p_3=0$ consists of a sheet ${\rm Ro}=-(1+{\rm Lu}^2)^{-1}$
corresponding to the doubly degenerate infinite values of ${\rm Rm}$
at the maxima of the helically modified SMRI. It smoothly touches along
the $\beta$-axis the surface $\Delta=0$ that consists of two smooth sheets
that touch each other along a spatial curve --- the cuspidal edge ---
corresponding to triple roots of the polynomial $Q$,
Fig.~\ref{fig:9}(a).

Every point on the upper sheet of the surface $\Delta=0$
represents a degenerate linear approximation to the essential HMRI
domain and therefore a maximal $\rm Ro$ at the corresponding values of
$\beta$ and ${\rm Lu}$. Numerical optimization
results in the new ultimate limit for HMRI
${\rm Ro}_{\rm c} \simeq-0.802$ at ${{\rm Lu}}\simeq0.618$,
$\beta\simeq0.634$, and ${\rm Rm}\simeq0.770$, see Fig.~\ref{fig:9}(b).
This new limit of ${\rm Ro}_{\rm c}$ on the cuspidal edge is smoothly connected to the
inductionless Liu limit by
the upper sheet of the discriminant surface, which converges to the curve \rf{r2}
when ${\rm Lu}=0$.
We point out that the new limit is achieved at
${\rm Ha}\rightarrow \infty$ when the
optimal $\rm Pm$ tends to zero in such a way that
${{\rm Lu}}\simeq0.618$.
Figure ~\ref{fig:9}(c) shows the behaviour of the so-called
interaction parameter (or Elsasser number) $\rm N=Lu^2/Rm$ for the
HMRI sheet. It is remarkable that, at $\rm Lu=0$, HMRI starts to work
already at $\rm N=0$. This can be explained by the observation that
the optimal value for HMRI corresponds to
${\rm N}{\rm Ha}={\rm Lu}^3 /({\rm Rm}\sqrt{\rm Pm})=1/(1+2^{-1/2})=0.586$,
\cite{ks10}. Later, for increasing
$\rm Lu$, the optimal ${\rm N}$ acquires final values, passes through its maximum
and at ${{\rm Lu}}\simeq0.618$ and
$\beta\simeq0.634$ it terminates at $\rm N=0.496$.

\section{Conclusion}

Motivated by the well-established theory of dissipation induced instabilities  
\cite{Bo56,Ar71,Le80,Le82,GKL90,BKMR94,HR95,L03,K04,K07,KM07,KV10}, we
have resolved the two paradoxes of SMRI and HMRI in the
limits of infinite and zero magnetic Prandtl number, respectively,
by establishing their sharp correspondence to singularities
on the instability thresholds.
In either case, the local Pl\"ucker conoid structure has been identified
as responsible for the non-uniqueness of the critical Rossby number, and its
crucial dependence on the Lundquist number. For HMRI, we have found an
extension
of the former Liu limit ${\rm Ro}_{\rm c} \simeq -0.828$
(valid for ${\rm Lu}=0$) to a somewhat higher
value ${\rm Ro}\simeq-0.802$ at ${\rm Lu}=0.618$ which is, however,
still below the Kepler value.
A remarkable feature of HMRI is an abrupt disappearance of its extrema at a
finite Lundquist number (see Fig. 9). The discussion of possible     
physical consequences of this discontinuity, for example         
as an alternative way of  explaining the so-called          
Quasi-Periodic-oscillations (QPO) \cite{ReMc06,LeBaLa09},
must be left for future work.

\section*{Acknowledgments}
Financial support from the Alexander von Humboldt Foundation and the DFG
in frame of STE 991/1-1 and of SFB 609 is gratefully acknowledged.


\end{document}